\documentclass[twocolumn,aps,prl,showpacs,amsmath,amssymb,floatfix]{revtex4-1}
\usepackage{graphicx}
\usepackage{dcolumn}
\usepackage{bm}
\usepackage{xcolor}

\begin{document}

\title{Tuning contact line dynamics and deposition patterns in volatile liquid mixtures}

\author{Asher P. Mouat}
\author{Clay E. Wood}
\author{Justin E. Pye}
\author{Justin C. Burton}
\email{justin.c.burton@emory.edu}

\affiliation{Department of Physics, Emory University, Atlanta, Georgia 30322, USA}

\date{\today}

\begin{abstract}
The spreading of a pure, volatile liquid on a wettable substrate has been studied in extensive detail. Here we show that the addition of a miscible, non-volatile liquid can strongly alter the contact line dynamics and the final liquid deposition pattern. We observe two distinct regimes of behavior depending on the relative strength of solutal Marangoni forces and surface wetting. Finger-like instabilities precede the deposition of a sub-micron thick film for large Marangoni forces and small solute contact angles, whereas isolated, pearl-like drops emerge and are deposited in quasi-crystalline patterns for small Marangoni forces and large solute contact angles.  This behavior can be tuned by directly varying the contact angle of the solute liquid on the solid substrate.
\end{abstract}

\maketitle

The spreading and evaporation of a volatile liquid is pervasive in nature. Every raindrop splash left on a surface will end its existence by evaporation, leaving behind any dissolved contaminants. If the contaminants are colloidal particles, strong evaporation near the contact line leads to a fluid flow towards the edge of the drop, producing well-known coffee ring patterns \cite{Deegan1997,Deegan2000,Kaplan2015,Yunker2011,Kim2016}. 
However, much less is known about the deposition of a non-volatile liquid (solute) dissolved in a volatile liquid  (solvent). Pure, refined liquids are used throughout the natural sciences and engineering for controlled experiments, cleaning, and evaporation-assisted surface patterning technologies \cite{Prasad2004,Gonuguntla2004,Doumenc2013,Deblais2016}, yet they often contain residual liquid solutes from the manufacturing process. The deposition pattern will depend on the solute concentration, surface tension gradients (Marangoni effects), and the wetting properties of the substrate. Many studies have revealed novel, microscopic contact line instabilities resembling fingers driven primarily by thermal or surfactant-based Marangoni forces \cite{Troian1990,Cazabat1990,Sur2004,Gotkis2006,Wodlei2018}. Although less studied, solutal Marangoni forces are responsible for droplet ``bursting'' patterns on liquid surfaces \cite{Keiser2017}, can delay mixing of miscible liquids \cite{Kim2017}, and produce the well-known ``tears of wine'' phenomenon \cite{Fournier1992,Fanton1998,Hosoi2001,Venerus2015}.  In wine, the evaporation of the volatile solute (ethanol) leads to a higher surface tension near the contact line which pulls the liquid film up the sides of the glass. 

Here we show how contact-line instabilities driven by solvent evaporation in a spreading drop can determine the deposition pattern of a partially-wetting liquid solute on a surface. As the wetting solvent rapidly evaporates at the contact line, a thick rim of solute forms and breaks up into individual drops whose size decreases with the solute concentration. Further evolution of the drops is determined by a balance of solutal Marangoni forces and surface wetting forces. Strong Marangoni forces and surface wetting lead to the emergence of finger-like protrusions at the contact line and the deposition of a sub-micron thick film that remains after the solvent evaporates. If these forces are too weak, the drops grow into bulbous ``pearls'' that are often deposited in striking, quasi-crystalline patterns.  We also show how a reduction of the solute's equilibrium contact angle below a threshold can demarcate these vastly different behaviors, and provide a quantitative estimate of this threshold that only depends on the surface tensions of the liquids. 

\begin{figure*}
\includegraphics[width=6.4 in]{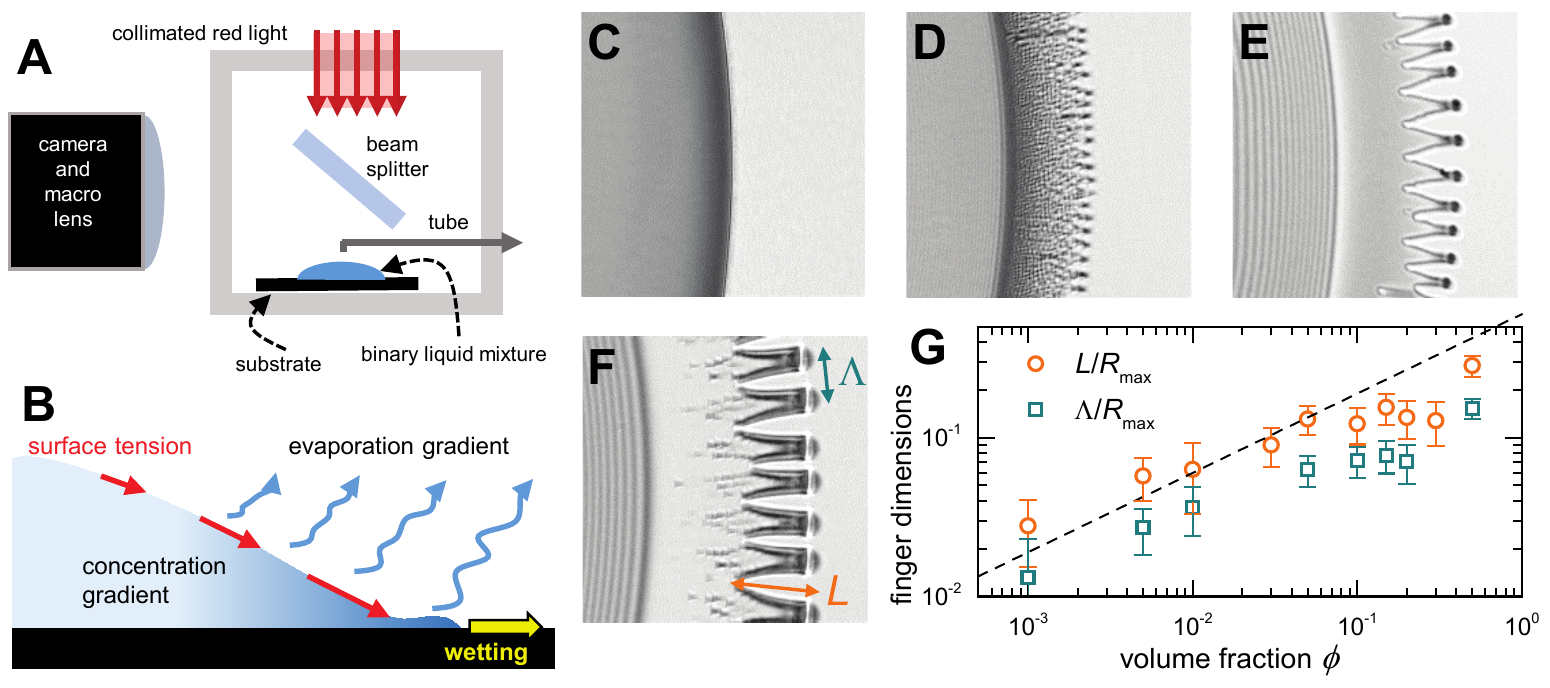}
\caption[]{(\textbf{A}) Experimental setup. A drop is deposited on a silicon wafer and imaged with monochromatic light as it spreads. (\textbf{B}) Evaporation induces concentration and surface tension gradients within the spreading drop. Images show the contact line near $R_{\text{max}}$ for binary mixtures of isopropanol with ethylene glycol at $\phi$ = 0 (\textbf{C}), 0.001 (\textbf{D}), 0.005 (\textbf{E}), and 0.01 (\textbf{F}). All images are 1 mm $\times$ 1 mm. (\textbf{F}) Finger length and wavelength vs. solute volume fraction. The dashed line represents $\phi^{1/2}$. 
\label{fig_1}} 
\end{figure*}

Our experiments consisted of quantitative, interferometric imaging of spreading drops on smooth, oxidized silicon wafers \cite{Poulard2003,Burton2012}. A diagram of the setup is shown in Fig.\ \ref{fig_1}A. Monochromatic red light from a solid state source ($\lambda$ = 632 nm, coherence length $\approx$ 10 $\mu$m) was passed through a 50-50 aluminized beam splitter. The spreading drops were imaged with a digital camera at a resolution of 6 $\mu$m/pixel. Reflections from the liquid-air interface and the silicon wafer produced observable patterns of interference fringes. Drops of volume $V_\text{0}$ = 1 $\mu$l with solute volume fraction $\phi$ were deposited by a syringe pump onto oxidized silicon wafers in a closed environment at 22$^{\circ}$C and $45\pm5$\% relative humidity. The silicon wafers were ultrasonically cleaned with deionized water and $>$99.9 vol\% pure isopropanol, then dried with nitrogen gas and stored in a clean oven prior to use. All liquids were purchased from Fisher Scientific with $>$99 vol\% purity. For some experiments, the wafer was treated with oxygen plasma for 30-60 s in a custom-built apparatus based on a consumer-grade microwave oven. Surface treatment with highly-reactive oxygen plasma removes organic contaminants and generates functional hydroxyl groups on the SiO$_2$ surface layer. The result is a dramatic increase in the hydrophilicity of polar liquids such as water, and a reduced contact angle. 

When a liquid spreads on a thermally-conducting surface, the evaporation flux is highest near the contact line since heat can be rapidly delivered to the liquid-vapor interface \cite{Deegan2000,Kavehpour2002,Ristenpart2007,Karpitschka2017}. For drops composed of a non-volatile liquid solute (low vapor pressure $p_\text{v}$) dissolved in a volatile liquid (high $p_\text{v}$), this evaporation induces a positive solute concentration gradient (Fig.\ \ref{fig_1}B). If the solute's liquid-vapor surface tension ($\gamma^{\text{lv}}_2$) is larger than the solvent's ($\gamma^{\text{lv}}_1$), a tangential Marangoni stress will act at the interface in the radial direction, in addition to wetting forces at the contact line. In our experiments, we measured the equilibrium contact angle, $\theta_{\text{eq}}$, of each liquid by optical imaging. Measurements of $\gamma^{\text{lv}}$ and $p_\text{v}$ for pure fluids were taken from Refs. \cite{Jasper1972,CRC2018}, and the surface tensions of solute mixtures were measured using axisymmetric drop shape analysis \cite{Burton2010}. The relevant properties ($\gamma$ [mN/m], $p_\text{v}$ [Pa], $\theta_\text{eq}$ [$^\circ$]) for each liquid were: isopropanol (21.5, 5000, 0), ethylene glycol (48.0, 13, 30), propylene glycol (36.6, 17, 22), dodecane (25.0, 14, 5), water (72.0, 2530, 44), dimethyl sulfoxide (43.5, 30, 30), and glycerol (64.0, 0.022, 44). 

For pure liquids spreading on clean, silicon wafers, instabilities were not observed at the contact line (Fig.\ \ref{fig_1}C). This contrasts with results shown in Gotkis et al. \cite{Gotkis2006} for isopropanol on silicon. The authors reported finger-like instabilities at the contact line resembling ``octopi'' and measuring over 100 $\mu$m in length. These instabilities were attributed to thermal Marangoni forces. However, for thin liquid films on substrates with a high thermal diffusivity (i.e. silicon), we estimate that thermal effects are insufficient to initiate such instabilities (see SI \cite{supp}). Instead, we found that a small amount of solute could easily produce finger-like instabilities. Figure \ref{fig_1}D shows the spreading of an isopropanol drop with ethylene glycol at $\phi=0.001$. Small drops of the ethylene glycol are jettisoned in front of the main drop due to solutal Marangoni forces, then deposited on the surface, and remain after evaporation of the solvent (video S1 \cite{supp}). 

For higher concentrations of the solute liquid, well-defined fingers formed that were attached to a sub-micron thick film. Figure \ref{fig_1}E-F shows the spreading of isopropanol drops with ethylene glycol at $\phi=0.005$ and $\phi=0.01$ (video S2 \cite{supp}). The fingers were preceded by a thick rim of solute that developed from solvent evaporation near the contact line. Our observations suggest that this rim breaks up due to the Rayleigh-Plateau instability \cite{Wodlei2018,Eggers1997} (Fig.\ S1 \cite{supp}), and the resulting drops act as progenitors to the emerging fingers. This is in agreement with recent results for the wavelength of tears of wine \cite{Venerus2015}. Additionally, both the finger length, $L$, and finger wavelength, $\Lambda$, decrease with $\phi$, and are consistent with the scaling $\phi^{1/2}$ for $\phi\lesssim0.1$ (Fig.\ \ref{fig_1}G). We can approximate the volume of the concentrated solute region as a thin torus of radius $R$ and thickness $a$ with volume $V_0 \phi\propto R a^2$. The fingers are comprised mostly of solute, thus their characteristic size will be $a\propto\phi^{1/2}$. The full evolution of the fingers will also depend on the local film thickness, similar to other fingering instabilities described in driven, spreading liquid films \cite{Huppert1982,Fournier1992,Troian1989,Jerrett1992,Oron1997,Kondic1999,Kondic2003,Grigoriev2003,Sur2004}.

\begin{figure}
\begin{center}
\includegraphics[width=3.2 in]{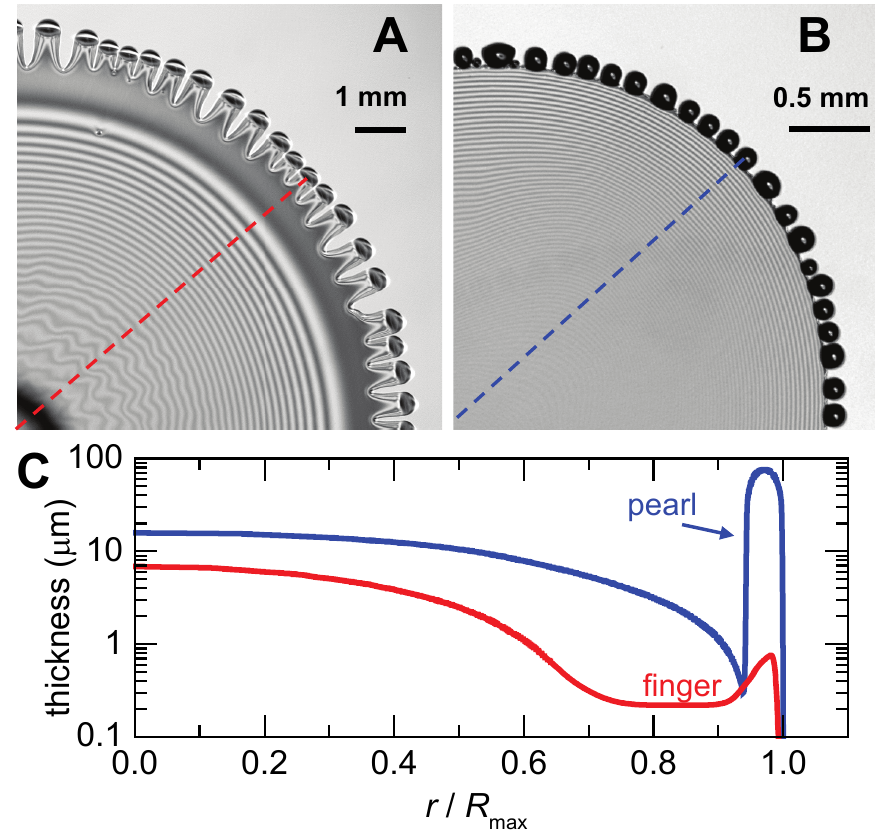}
\caption[]{Partial image of an isopropanol drop with ethylene glycol ($\phi$ = 0.1, \textbf{A}) and dodecane ($\phi$ = 0.1, \textbf{B}) spreading on a clean silicon wafer. The thickness difference between each black and white fringe is $\approx$ 113 nm. By counting fringes along a bisect in an image, such as the dashed red line in \textbf{A}, we can obtain the thickness profile of the drop (\textbf{C}), up to the highly-curved finger tips (we assume the drops terminate at the surface). 
\label{fig_2} }
\end{center} 
\end{figure}

We can reduce the influence of Marangoni forces by using a solute with a surface tension comparable to the solvent. In this regime we observe round ``pearls'' instead of elongated fingers.  
Figure \ref{fig_2}A-B shows images of isopropanol drops spreading with ethylene glycol ($\phi=0.1$), and dodecane ($\phi=0.1$), respectively. The surface tension gradient between isopropanol and dodecane is $\approx$ 10 times smaller than for ethylene glycol. Since the optical indices of all liquids in the experiments ($1.33<n<1.47$) were smaller than silicon at $\lambda$ = 632 nm, the first destructive interference fringe corresponded to a thickness = $\lambda/4n\approx$ 113 nm for $n\approx$ 1.4. Thinner regions are essentially transparent. The uniform intensity in the thin film surrounding the central part of the drop in Fig.\ \ref{fig_2}A indicates that the thickness is nearly uniform. We obtained measurements of the absolute thickness of the ethylene glycol film by observing the final moments of evaporation and counting changes in fringe intensity backwards in time (Fig.\ \ref{fig_2}C). For dodecane, the film decreased smoothly until it abruptly ended in a bulbous pearl (Fig.\ \ref{fig_2}B), which was assumed to be a hemisphere attached to the film on one side so that its thickness could be determined.  



\begin{figure}
\begin{center}
\includegraphics[width=3.2 in]{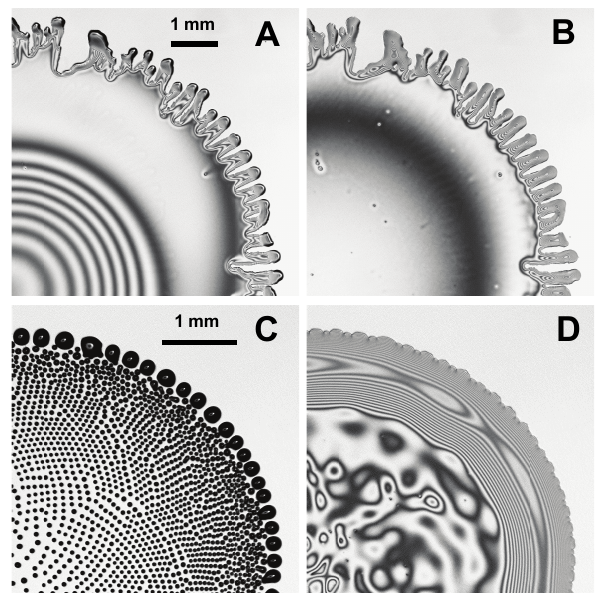}
\caption[]{(\textbf{A-B}) Partial images of an isopropanol drop with propylene glycol at $\phi=0.1$. The time elapsed between the 2 images is $\approx$ 25 s. A thin, uniform film of liquid remains after evaporation of the isopropanol. (\textbf{C-D}) Partial images of an isopropanol drop with dodecane at $\phi=0.1$. The time elapsed between the 2 images is $\approx$ 32 s. The droplets eventually coalesce into a continuous film. The scale bar applies to all images in a sequence.
\label{fig_3}} 
\end{center} 
\end{figure}

The emergence of fingers and the trailing thin film determined the final deposition pattern of the solute. Figure \ref{fig_3}A-B shows images from an isopropanol drop with propylene glycol at $\phi=0.1$. The thin, uniform film seen in \ref{fig_3}B was drawn out by the fingers and eventually evaporated. In contrast, the formation of pearls at the contact line was associated with droplet deposition. Figure \ref{fig_3}C-D shows images from an isopropanol drop with dodecane at $\phi=0.1$ (video S4 \cite{supp}). Large drops of dodecane were deposited by the receding contact line, often leading to a quasi-crystalline pattern. The size of the drops decreased with $\phi$ (Fig.\ S2 \cite{supp}), and concentric rings of patterns were often produced (Fig.\ S3 \cite{supp}). Similar patterns have been studied in the context of colloidal and polymer deposition \cite{Kaplan2015,Larson2014,Gonuguntla2004,Wu2018,Lin2005}, as well as the dewetting of polymer films on surfaces \cite{Reiter2001} and liquid films on a liquid bath \cite{Keiser2017}. Long-ranged surface forces are also important for the final deposition pattern. Deposited films of ethylene glycol required a short, 10 s treatment with oxygen plasma to prevent retraction at long times (video S3 \cite{supp}). In addition, freshly-deposited drops of dodecane remained spherical (partial wetting) due to preferential wetting of isopropanol on the surface (Fig.\ \ref{fig_3}C) \cite{supp}. After the isopropanol diffused away from the surface, the dodecane droplets coalesced into a thick film (Fig.\ \ref{fig_3}D).

\begin{figure}
\begin{center}
\includegraphics[width=3.2 in]{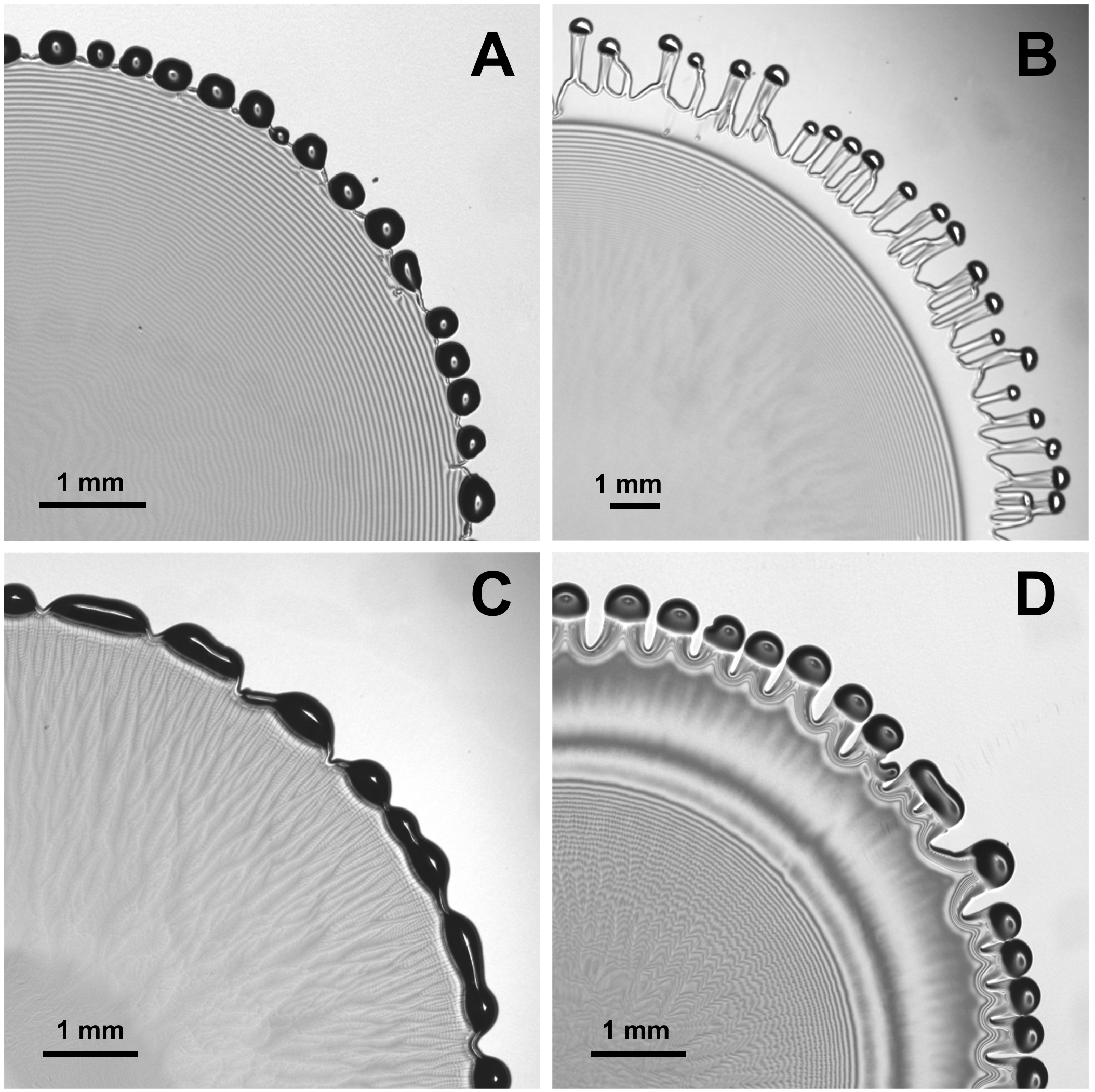}
\caption[]{Images showing the spreading of an isopropanol drop with water at $\phi$ = 0.25 on a clean silicon substrate (\textbf{A}) and on a surface treated with oxygen plasma for 30 s (\textbf{B}). Images are also shown for an isopropanol drop with glycerol at $\phi$ = 0.2 on a clean silicon substrate (\textbf{C}) and on a surface treated with oxygen plasma for 30 s (\textbf{D}). Note the difference in scale bars. 
\label{fig_4}} 
\end{center} 
\end{figure}

One may expect solutes such as water to readily form fingers due to their large surface tension, however, strong Marangoni forces alone are not sufficient. Rather, on a clean silicon surface, water formed well-defined pearls, as shown in Fig.\ \ref{fig_4}A, and also reported in Ref.\ \cite{Gotkis2006}. The lack of fingers is due to water's weak affinity for the clean silicon surface ($\theta_\text{eq}\approx 44^{\circ}$). We confirmed this by treating the surface with oxygen plasma for 30 s, which resulted in $\theta_{\text{eq}}\approx 10^{\circ}$, the emergence of fingers, and an eventual sub-micron thin film (Fig.\ \ref{fig_4}B). However, further treatment with oxygen plasma suppressed wetting for isopropanol. In this case, neither fingers nor pearls formed (video S5 \cite{supp}). The robust interplay between Marangoni and surface wetting forces was present even for high-viscosity solutes. Figure \ref{fig_4}C shows the spreading of an isopropanol drop with glycerol at $\phi$ = 0.2, where the solute viscosity (1180 mPa.s) is 2 orders of magnitude larger than most fluids in our experiments. Pearls form at the contact line, and upon evaporation, isolated glycerol drops are left behind that do not wet the surface (Fig.\ S3). In contrast, treating the surface with oxygen plasma lead to well-defined fingers (Fig.\ \ref{fig_4}D) and a thin residual film of glycerol. Due to glycerol's hygroscopic properties and sharp variation of viscosity with water content, we rinsed the oxygen plasma-treated slide with deionized water and dried it prior to deposition to obtain reproducible results. 

\begin{figure}
\begin{center}
\includegraphics[width=3.4 in]{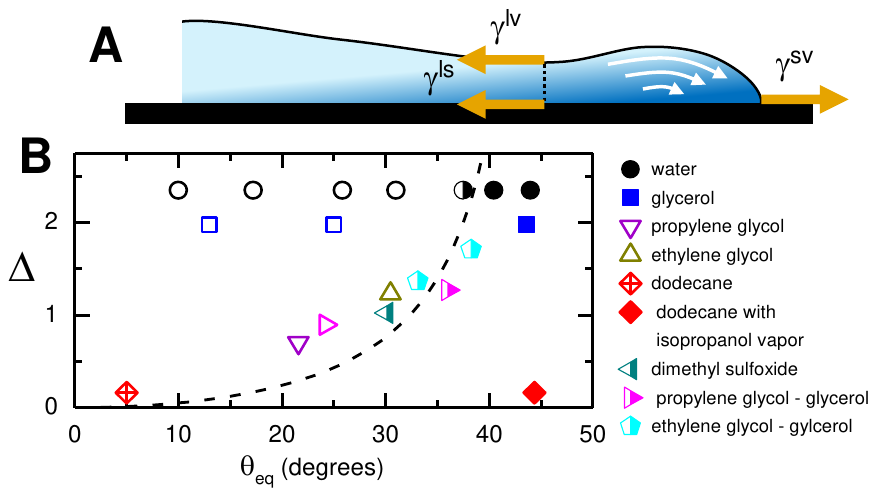}
\caption[]{ (\textbf{A}) A solute rim, delineated by the black dotted line, will extend provided that the net surface tension pulls the rim outward. White arrows indicate flow. (\textbf{B}) Phase portrait of deposition behaviors versus $\Delta$ and $\theta_\text{eq}$. Open symbols correspond to fingers and thin film deposition, closed symbols correspond to pearls and drop deposition. Half-filled symbols represent a mixture of both behaviors (small fingers connected to pearls). Multiple points for water and glycerol represent different waiting times after the initial plasma cleaning. The dashed line is given by Eq.\ \ref{fbal} with $\alpha=0.69$.
\label{fig_5}} 
\end{center} 
\end{figure}

The appearance of fingers or pearls depended most sensitively on the difference in surface tension and wetting properties of the liquids. Thus, we can construct a quantitative estimate of the boundary separating the qualitative deposition patterns.  Figure \ref{fig_5}A shows a cross section of a newly-emerged rim at the contact line and the relevant surface tension forces in the radial direction. For simplicity, we ignore dynamic pressure and viscous drag forces in the thin liquid film, and assume that the solute gradient is strongest near the liquid-vapor surface where evaporation is occurring. At the liquid-vapor interface, $\gamma^\text{lv}\approx(1-\alpha)\gamma^\text{lv}_1+\alpha\gamma^\text{lv}_2$, where l, s, and v refer to the liquid, solid, and vapor phases, and the subscript refers to the solute (2) or solvent phase (1). The parameter $\alpha$ is the local volume fraction of solute. In the absence of slip near the substrate, the concentration of the solute will be high since the clean surface is coated by fluid at the contact line, thus $\gamma^\text{ls}\approx \gamma^\text{ls}_2$. The solid-vapor surface tension, $\gamma^\text{sv}$, pulls the contact line outward. 


The boundary between the formation of fingers and pearls can found by equating the forces  and making use of the Young-Dupr\'{e} equation, $\gamma_2^\text{lv}\cos(\theta_\text{eq})=\gamma^{\text{sv}}-\gamma_2^{\text{ls}}$. The result is:
\begin{equation}
\dfrac{\gamma_2^\text{lv}-\gamma_1^\text{lv}}{\gamma_1^\text{lv}}=\Delta=\dfrac{1-\cos(\theta_\text{eq})}{\cos(\theta_\text{eq})-\alpha}.
\label{fbal}
\end{equation}
Figure \ref{fig_5}B shows a phase portrait of the deposition patterns using $\Delta$ and $\theta_\text{eq}$. In addition to surface plasma treatment, we also used glycol-glycerol solute mixtures in order to tune between the deposition patterns. For dodecane, we included the value for $\theta_\text{eq}$ on clean silicon and in the presence of a saturated isopropanol atmosphere (where $\theta_\text{eq}\approx 44^\circ$). Equation\ \ref{fbal} shows excellent agreement with the experimental data using $\alpha=0.69$. We note that this boundary applies to a wide range of volume fractions tested in our experiments, $0.001<\phi<0.3$. Higher concentrations of solute ($\phi>0.5$) are related to ``tears of wine'', where the non-volatile component is dominant, and gravitational forces are important \cite{Fournier1992,Fanton1998,Hosoi2001,Venerus2015}. 

In conclusion, we have shown how the contact line dynamics and deposition pattern of one miscible liquid in a volatile solvent has two distinct regimes characterized by surface wetting and solutal Marangoni forces. Low contact angles and large Marangoni forces lead to the emergence of fingers and a persistent, sub-micron thick film, whereas large contact angles and small Marangoni forces lead to pearls and the deposition of isolated drops. The boundary between regimes does not depend on solute concentration or solute viscosity. Although the phase portrait shown in Fig.\ \ref{fig_5}B is specific to our solvent (isopropanol), Eq.\ \ref{fbal} is quite general, and we have seen analogous behavior with other solvents such as acetone. Thus, we expect the qualitative boundary between the two regimes will remain provided the volatile solvent wets the surface under investigation. We also note that these results may provide a low-cost method for making large ($\sim$ cm$^2$) areas of microscopic liquid films for colloidal particle and macromolecule deposition on surfaces. However, these investigations are left for future studies. 

\begin{acknowledgments} 
This work was supported by the NSF DMR Grant No. 1455086. 
\end{acknowledgments}

\end{document}